\documentclass[12pt,preprint]{aastex}

\usepackage{natbib}
\usepackage{enumerate}
\usepackage{enumitem}

\slugcomment{}

\shorttitle{}
\shortauthors{Chen et al.}

\begin{document}


\title{Tether-cutting Reconnection between Two Solar Filaments Triggering Outflows and a Coronal Mass Ejection}

\author{Huadong Chen\altaffilmark{1}, Jun Zhang\altaffilmark{1}, Leping Li\altaffilmark{1}, 
Suli Ma\altaffilmark{2}}
\email{hdchen@nao.cas.cn}
\altaffiltext{1}{Key Laboratory of Solar Activity,
     National Astronomical Observatories, Chinese Academy of Sciences,
      Beijing 100012, China}
\altaffiltext{2}{College of Science, China University of Petroleum, Qingdao 266580, China}



\begin{abstract}
Triggering mechanisms of solar eruptions have long been a challenge. 
A few previous case studies have indicated that preceding gentle filament merging via magnetic reconnection may launch following intense eruption, according with the tether-cutting (TC) model.
However, detailed process of TC reconnection between filaments has not been exhibited yet.
In this work, we report the high resolution observations from the $Interface\ Region\ Imaging\ Spectrometer$ ($IRIS$) of TC reconnection between two sheared filaments in NOAA active region 12146.
The TC reconnection commenced since $\sim$15:35 UT on 2014 August 29 and triggered an eruptive $GOES$ C4.3-class flare $\sim$8 minutes later.
An associated coronal mass ejection appeared in the field of view of $SOHO$/LASCO C2 about 40 minutes later.
Thanks to the high spatial resolution of $IRIS$ data, bright plasma outflows generated by the TC reconnection are clearly observed, which moved along the subarcsecond fine-scale flux tube structures in the erupting filament.
Based on the imaging and spectral observations, the mean plane-of-sky and line-of-sight velocities of the TC reconnection outflows are separately measured to be $\sim$79 and 86 km s$^{-1}$, which derives an average real speed of $\sim$120 km s$^{-1}$.
In addition, it is found that spectral features, such as peak intensities, Doppler shifts, and line widths in the TC reconnection region evidently enhanced compared with those in the nearby region just before the flare.
\end{abstract}





\section{Introduction}
Solar eruptions including filament or prominence eruptions, flares, and coronal mass ejections (CMEs) are commonly believed to be related to magnetic reconnection, during which magnetic energy can be converted into thermal and kinetic energy of the plasma \citep[e.g.,][]{lin00, priest14}. 
Some typical theoretic models have been devoted to the interpretations of eruption initiations, such as the tether-cutting (TC) or flux cancellation model \citep{moore01, kusano04, kusano12, van89, fletcher01, joshi15, zhang15}, magnetic breakout model \citep{antiochos99}, emerging flux mechanism \citep[e.g.,][]{chen00, louis15}, and ideal magnetohydrodynamic (MHD) instabilities \citep[e.g.,][]{torok04, kliem06}.
In the TC model proposed by \citet{moore01}, magnetic reconnection in the lower solar atmosphere is considered for the destabilization of the system, which takes place between the inner legs of the sheared core fields in an active region (AR).
The MHD simulations performed by \citet{kusano12} further show that there are two possible processes triggering flare through the TC reconnection, named ``eruption-induced reconnection" and ``reconnection-induced eruption".

TC model has been supported by some observational phenomena, such as the brightenings in multi-wavebands \citep[e.g.,][]{yurchyshyn06}, slow-rise motion of filaments \citep[e.g.,][]{sterling11}, and morphological changes of flaring structures \citep[e.g.,][]{cheng14}.
Recently, Chen et al. (2014) presented an unambiguous observation of TC reconnection, which occurred between sheared coronal loops in an AR and triggered a halo CME. 
Chen et al. (2015) found that at least four homologous confined X-flares in NOAA AR 12192 were initiated by TC reconnections.
Filament interactions or mergings via magnetic reconnection have been reported in some observational studies \citep[e.g.,][]{schmieder04, su07, kumar10, bi12, liy12, liu12, kong13, jiang14, zhu15} and modeled in numerical works \citep[e.g.,][]{linton01, devore05, aulanier06, torok11}.
Most of the observations show that the filament interaction takes place when the filaments collide with each other during their eruption, reflecting the filament interaction results from the eruption.
While in a few cases \citep[e.g.,][]{bone09, joshi14}, the filaments merge with each other firstly and then give rise to the ensuing eruption, in good agreement with the TC model.

In this study, using the high-resolution data from the $Interface\ Region\ Imaging\ Spectrometer$ \citep[$IRIS$;][]{depontieu14} and $Solar\ Dynamics\ Observatory$ \citep[$SDO$;][]{pesnell12}, we present a distinct observation of TC reconnection during an eruption event, which occurred between two sheared filament-carrying magnetic fields in NOAA AR 12146 and triggered a subsequent $GOES$ C-class flare and a CME. 
By virtue of the high spatial resolution of $IRIS$ slit jaw imager (SJI) data, outflows from the TC reconnection region along subarcsecond flux tube structures are firstly unveiled clearly.
Combined with $IRIS$ spectroscopic observations, the kinematics of reconnection outflows and spectral characteristics of the TC reconnection region are investigated as well.


\section{Observations and Data Analysis}
A solar eruption occurred at the western solar limb on 2014 August 29, starting at $\sim$15:35 UT. 
This event was observed by $IRIS$ with the 1330 \AA\ SJI using a spatial scale of $\sim$0\arcsec.166 and a cadence of  $\sim$9.6 s.
The $IRIS$ spectral data are taken in a large coarse 8-step raster mode with a 76 s cadence and a spectral resolution of $\sim$0.025 \AA.
The eight positions of the scanning spectrometer slit (S1--S8) are indicated by the discontinuous vertical lines in Figure~1(c).
We mainly analyzed the emission of the \ion{Si}{4} 1393.76 \AA\ line, which is formed in the middle transition region with a temperature of $\sim$0.08 MK \citep{peter14}.
The nearby \ion{S}{1} 1392.589 \AA\ and \ion{Fe}{2} 1392.816 \AA\ lines are used for the absolute wavelength calibration \citep{tian15}.
We use the level 2 data, which have been dark corrected, flat fielded, and geometrically corrected.
A double-Gaussian fit to the \ion{Si}{4} 1394 \AA\ line profile is performed to study the outflow kinematics.
For a rough analysis to the spectra of the TC reconnection region, we also apply a single-Gaussian fit to the \ion{Si}{4} line \citep{peter14} and obtained the temporal evolutions of peak intensity, Doppler shift, and line width (see Figure~4).
The multi-wavelength intensity images and longitudinal magnetograms from the Atmospheric Imaging Assembly \citep[AIA;][]{lemen12} and Helioseismic and Magnetic Imager \citep[HMI;][]{schou12} on-board $SDO$ are also utilized. 

\section{Results}
\subsection{Filaments Eruption Triggered by TC Reconnection} \label{}
According to $GOES$ soft X-ray (SXR) observations, a C4.3-class flare associated with the eruption took place in AR 12146 ($\sim$N06W90) from 15:43 UT on August 29. 
About 30 minutes later, a CME appeared in the field of view (FOV) of the Large Angle and Spectrometric Coronagraph \citep[LASCO;][]{brueckner95} C2 and spread outwards with a median velocity of $\sim$312 km s$^{-1}$ \citep[see CACTus catalogue;][]{robbrecht09}.
According to the high spatial-resolution $IRIS$ 1330 \AA\ SJI data, it can be seen that there exist two different filaments F1 and F2 before the eruption.
Their spines are outlined by the dotted curves in Figure~1(a).
To further confirm this conclusion, we checked the AIA 304 \AA\ images one day before and found that the two filaments indeed existed in AR 12146 on August 28 (see Figure~1(h)).
Note that SP1 and SP2 are two main spots of AR 12146.
Apparently, F1 and F2 are different filaments located in their respective channel at the AR southeastern periphery.
To avoid strong projection effect, we display one HMI longitudinal magnetogram of AR 12146 on August 27 in Figure~1(i).
Taking the spots SP1 and SP2 and some plages at the AR periphery as references, the approximate positions and magnetic polarities of the photospheric footpoints of F1 and F2 are inferred and marked with the red and green signs in Figure~1(i), respectively.
Since 09:05 UT on August 28, F1 and F2 only underwent some activations or disturbances partially and did not erupt until the onset of the event under study (see animation~2).

Summary images of the eruption observed by $IRIS$ 1330 \AA\ SJI are shown in Figures~1(a)--(f). 
More detailed 1330 \AA\ SJI observations (animation~1) reveal that F1 and F2 experienced a $\sim$1.5 hr slow-rise phase prior to impulsive acceleration, which is consistent with the initiation phase of a CME reported by \citet{zhangjie01} and other similar observations in the extreme-ultraviolet (EUV) wavebands \citep[e.g.,][]{sterling11, chen14}.
From 15:35 UT, a sudden brightening appeared at the crossing position between the inner legs of F1 and F2 (see Figure~1(b)).
Then, probably due to the heating and accelerating by magnetic reconnection, some brightened plasma flowed out from the brightening region.
Meanwhile, a large-scale filament F with a dip structure formed and rose gradually (see Figure~1(d)).
It can be clearly seen that the mass flows moved outwards from the dip region along the two sides of F.
As F ascended continually, the associated flare was triggered and F started to erupt rapidly.
Figure~1(e) displays the complex twisted structure of the erupting F, which might be created during the reconnection \citep{priest14}.
In Figure~1(f), the flare ribbon near the observer is detected in the vicinity of the solar limb.
According to its spatial relationship with F and the classical two-ribbon flare scenario, two flare ribbons in this eruption are estimated to be located at the opposite-polarity magnetic field regions on the two sides of F (see Figure~5(c)).

In all AIA EUV channels, the sudden brightening and outflows were also observed, meaning a multi-temperature distribution from $\sim$10$^5$ to 10$^7$ K of the plasma in those regions.
Figure~1(g) provides the corresponding observation in the AIA 131 \AA\ channel.
This situation indicates that the brightening and outflows may be caused by magnetic reconnection.
In our observations, the slow rises of the filaments, sudden brightening at a low altitude, and newly formed large-scale erupting filament prior to the flare are well consistent with the descriptions of the TC model \citep{moore01}.
It is very likely that the TC reconnection between F1 and F2 destroys the magnetic balance of the field system and leads to the eruption. 

\subsection{Kinematics of the TC Reconnection Outflows}
The close-ups of the TC reconnection region in the SJI 1330 \AA\ channel are presented in the left column of Figure~2.
Several fine structures appearing as bright slim flux tubes (outlined by the dotted curves in Figures~2(a)--(c)) are clearly exhibited in the vicinity of the TC reconnection region (indicated by the ellipses).
According to our measurements, most of the thicknesses of these tubes are less than 1.$\arcsec$0, one of which is displayed by the yellow arrows in Figure~2(c).
Some large-scale flux tubes (blue) extend to the ends of the newly-formed F. 
While some small-scale tubes (purple) arise at the lower altitudes, which might be the manifestation of small flare loop generated by the TC reconnection \citep{chen14}.
As the TC reconnection evolved, the bright plasma can be seen to flow out from the reconnection region along these tubes.
The positions of the spectrometer slit S6, S8, and S1 in Figures~2(a)--(c) are near, to the right and left of the TC reconnection region, respectively.
Figures~2(d)--(f) display the corresponding \ion{Si}{4} 1394 \AA\ spectra from the three slit positions, which present some significantly blue- and redshifted features.
We chose six features (f1--f6 in Figures~2(d)--(f)) with relatively strong blueshifts or redshifts and found that all of them originated from the positions where the slit intersect with the flux tubes (marked with the pluses in Figures~2(a)--(c)).

Figures~3(a)--(f) show the \ion{Si}{4} 1394 \AA\ line profiles of f1--f6, respectively.
The non-axisymmetrical property of these profiles indicates the single-Gaussian fitting is not accurate for them.
Thus, we use a double-Gaussian function to fit the six spectra, respectively.
In Figures~3(a)--(f), the fitting results are indicated by the black curves, consisting of two single Gaussian components respectively denoted by the blue and red dotted profiles.
It can be seen that the fitting results are in good agreements with the observational spectra (pluses).
The peak intensities ($i$), Doppler shifts ($v$), and line widths ($w$) of the two fitting components are separately given by the blue and red values in Figures~3(a)--(f).
Apparently, the absolute value of $v_{2}$ is much larger than that of $v_{1}$.
According to the previous studies \citep[e.g.,][]{peter10}, the two Doppler shift components in our results are probably caused by two kinds of motions happened simultaneously: the movement of the flux-tube structure and the plasma flow along the tube.
Since the speed of the reconnection outflow might be larger than that of the flux tube, $v_{1}$ and $v_{2}$ should correspond to the line of sight (LOS) velocities of the flux tube and outflow, respectively.
According to our results, $v_{1}$ varies in the range [-0.6, 23.9] km s$^{-1}$, indicating most of these tubes moved far away from the observer; while $v_{2}$ changes from -136.2 to 110.8 km s$^{-1}$, implying the complex spatial orientations of these tubes.
The mean values of the unsigned $v_{1}$ and $v_{2}$ are 10.7 and 86.3 km s$^{-1}$, respectively.

To investigate the outflow motions in the plane of sky (POS), we made the time-distance maps from the SJI 1330 \AA\ images along two cuts (Cut1 in Figure~1(c) and Cut2 in Figure~2(b)) and show them in Figures~3(g) and (h), respectively.
The time-distance images clearly display the reconnection outflows from B to A along Cut1 or from D to C along Cut2.
Four tracks of the flows are picked and indicated by the dotted lines.
By using a linear fitting, we derived the POS velocities of the flows, which range from $\sim$58.0 to $\sim$100.7 km s$^{-1}$, with a mean of $\sim$79.0 km s$^{-1}$.
With regard to the flow with the POS velocity of 100.7 km s$^{-1}$ (see Figure~3(h)), its LOS velocity is $\sim$78.1 km s$^{-1}$ (see $v_{2}$ in Figure~3(e)).
Thus, the real velocity of the flow can be calculated to be about 127.4 km s$^{-1}$.
Taking the mean LOS velocity (86.3 km s$^{-1}$) and POS velocity (79.0 km s$^{-1}$) of the outflows into consideration, the average real velocity of the outflows is about 117.0 km s$^{-1}$.

\subsection{Spectral Characteristics of the TC Reconnection Region}
According to $IRIS$ spectral observations of this event, the TC reconnection region was mainly captured by the spectrometer slit at three scanning positions S5--S7 (see Figure~1(c)). 
To learn about the spectral characteristics of the TC reconnection region roughly, we applied a single-Gaussian fit to the spectra of the \ion{Si}{4} 1394 \AA\ line at each slit position and obtained the peak intensities, Doppler shifts, and line widths from the fitting results.
For a comparison, the temporal evolutions of the three parameters from 15:30 to 16:02 UT are displayed in the top (for S5--S7) and bottom (for S1--S3) rows of Figure~4.
We also plotted the $GOES$ SXR light curves in Figures~4(b) and (e).
As pointed out by the arrows and small windows in Figure~4, the intensities, Doppler shifts, and line widths in S5--S7 begin to enhance distinctly just before the associated C4.3 flare, while those in S1--S3 do not have prominent enhancements during the same period. 
According to our estimations, as for S5--S7, the mean values of the peak intensities (on a logarithmic scale), blueshifts, redshifts, and line widths in the area denoted by the small deformed window are $\sim$5.27, $\sim$$-$16.9 km s$^{-1}$, $\sim$15.7 km s$^{-1}$, and $\sim$42.2 km s$^{-1}$, respectively.
The corresponding mean values in S1--S3 are only $\sim$3.97, $\sim$$-$6.4 km s$^{-1}$, $\sim$8.2 km s$^{-1}$, and $\sim$19.9 km s$^{-1}$, respectively.
The line broadening in a precursor phase of solar flare has been pointed out by \citet{harra01, harra13}, while our results clearly reveal that the TC reconnection is a likely origin of this spectral feature.

\section{Summary and Discussion}
The possible scenario during the early phase of the event under study can be described by the schematic diagrams in Figure~5, which have backgrounds of an HMI continuum intensity image.
Prior to the beginning of the eruption, the filaments F1 and F2 were mainly situated at the polarity inversion line of the AR periphery. 
While the inside legs of F1 and F2 were brought together, magnetic reconnection took place and resulted in the appearances of small flare loop in the lower altitude and large-scale filament structure F connecting the far ends of F1 and F2.
In the meantime, the hot plasma heated and accelerated by the reconnection flowed out from the reconnection region along the reconnected field lines.
As the balance between the upward magnetic pressure and downward magnetic tension lost, the newly formed filament F together with the overlying background field commenced to erupt rapidly and led to the subsequent two-ribbon flare and CME.
Our observations presented here are in good agreement with the TC model suggested by \citet{moore01} and can be classified to the ``eruption-induced reconnection" event, according to \citet{kusano12}.

TC reconnections between filaments triggering eruptions have been reported in only a few cases.
For instance, \citet{bone09} gathered several sets of ground based H$\alpha$ data to report an increasingly dynamic process of two interacting filaments preceding the eruption;
\citet{joshi14} presented observations of compound flux rope formation via merging of two nearby filaments, which only led to a failed eruption.
Owing to the limitation of the instrument spatial resolution, the detailed processes of TC reconnections are not shown in their observations.
By virtue of the high resolution data from $IRIS$, we distinctly present a detailed process of TC reconnection between two sheared AR filaments, which apparently triggers the following flare and CME.
The subarcsecond field structures (flux tubes) near the TC reconnection region and outflows along these fine-scale structures prompted by the TC reconnection are clearly exhibited for the first time.

Coupled with $IRIS$ spectral data, the average real speed of the outflows is measured to be $\sim$120 km s$^{-1}$. 
Since the filaments (F1 and F2) are located at AR periphery and may be partially ionized, we assume the magnetic field strength of $\sim$20 G \citep{mackay10} and particle number density of $\sim$10$^{11}$ cm$^{-3}$ \citep{labrosse10} in the reconnection region, which derives a local Alfv$\acute{e}$n speed of $\sim$138 km s$^{-1}$.
The reconnection outflow speed we measured is basically in line with this value.
Recently, \citet{ning14} and \citet{reeves15} separately studied the reconnection outflows in a coronal bright point and a small prominence eruption event, which were observed along a small coronal loop and potentially reconnected field line, respectively.
A larger outflow velocity of $\sim$300 km s$^{-1}$ was found by them, implying the local field strengths and/or plasma environments between our event and theirs may be different.

\acknowledgments
$IRIS$ is a NASA small explorer mission developed and operated by LMSAL with mission operations executed at NASA Ames Research center and major contributions to downlink communications funded by ESA and the Norwegian Space Centre.
The $SDO$ data are courtesy of NASA, the $SDO$/AIA and HMI science teams.
This work was supported by NSFC (11533008, 41331068, 41204124, and 11221063), the Project Funded by China Postdoctoral Science Foundation (2015M571126), and the Strategic Priority Research Program---The Emergence of Cosmological Structures of the Chinese Academy of Sciences (No. XDB09000000).


\clearpage

\clearpage


\begin{figure}
\epsscale{1.0}
\plotone{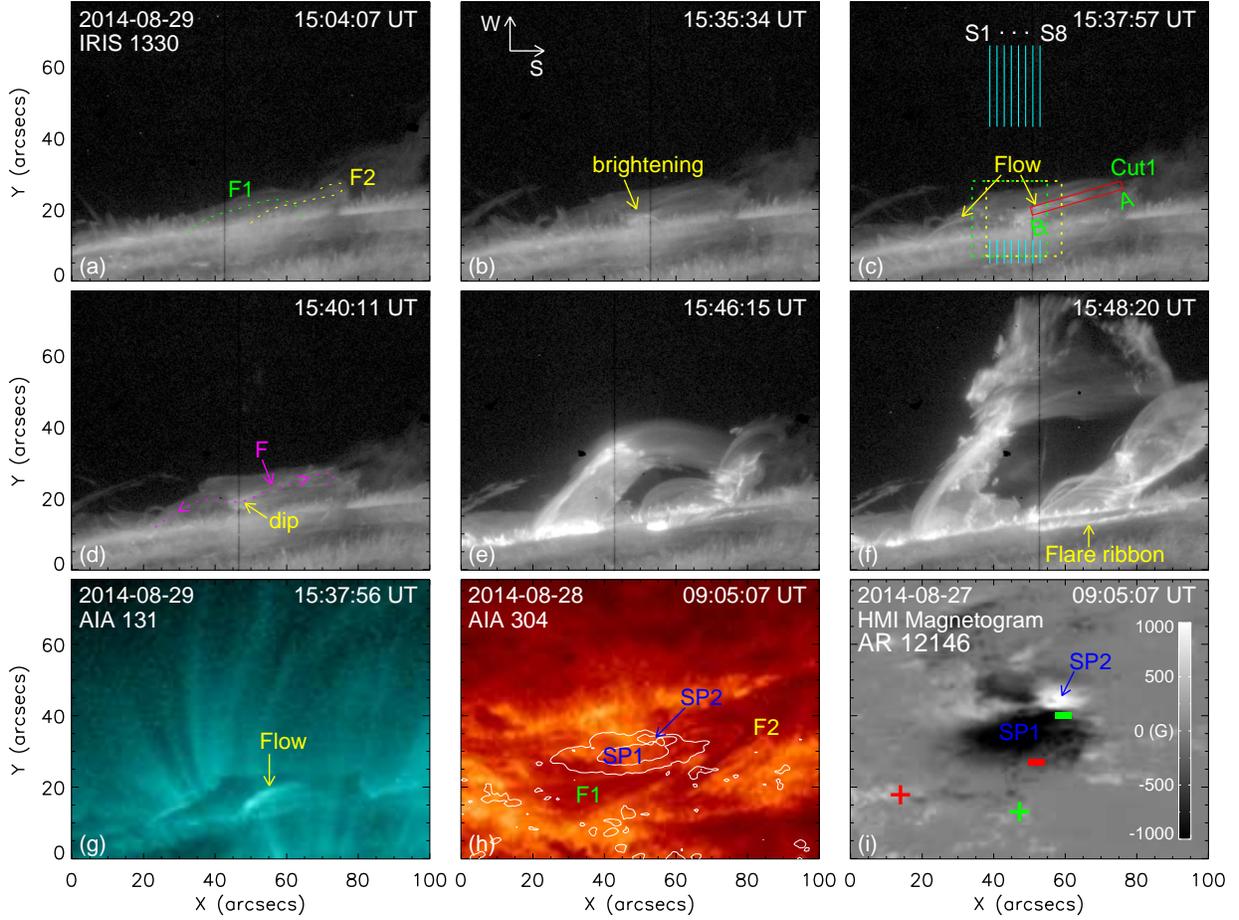}
\caption{((a)--(f)) $IRIS$ 1330 \AA\ SJI images showing the filament eruption 
(also see the animation~1);
((g)--(i)) AIA 131 \AA\ (g) and 304 \AA\ (h; also see the animation~2) images and HMI longitudinal magnetogram (i; also see animation~3).
The green and yellow dotted curves in panel (a) outline two different filaments F1 and F2.
the yellow and green dotted boxes indicate the FOVs of Figures~2(a)--(b) and Figure~2(c), respectively;
the red slim box indicate the cut used to make the time-distance map in Figure~3(g).
In panel (d), the purple dotted curve represents the outline of newly formed erupting filament F;
the arrows on the curve mark the outflow directions from the TC reconnection region.
The white contours in panel (h) are from the HMI continuum intensity map.
The contour levels are 46\%, 67\%, and 88\% of the maximum value, respectively.
SP1 and SP2 in panels (h) and (i) denote two main spots in AR 12146.
The red and green signs in panel (i) mark the possible positions of the footpoints of filament F1 and F2, respectively.
The $IRIS$, AIA and HMI images have been rotated counterclockwise by 90\degr.
The centers of panels (a)--(g), (h), and (i) are at solar (x,y) = (971\arcsec,120\arcsec), (924\arcsec,95\arcsec), and (791\arcsec, 66\arcsec), respectively.
\label{fig1}}
\end{figure}

\begin{figure}
\epsscale{0.7}
\plotone{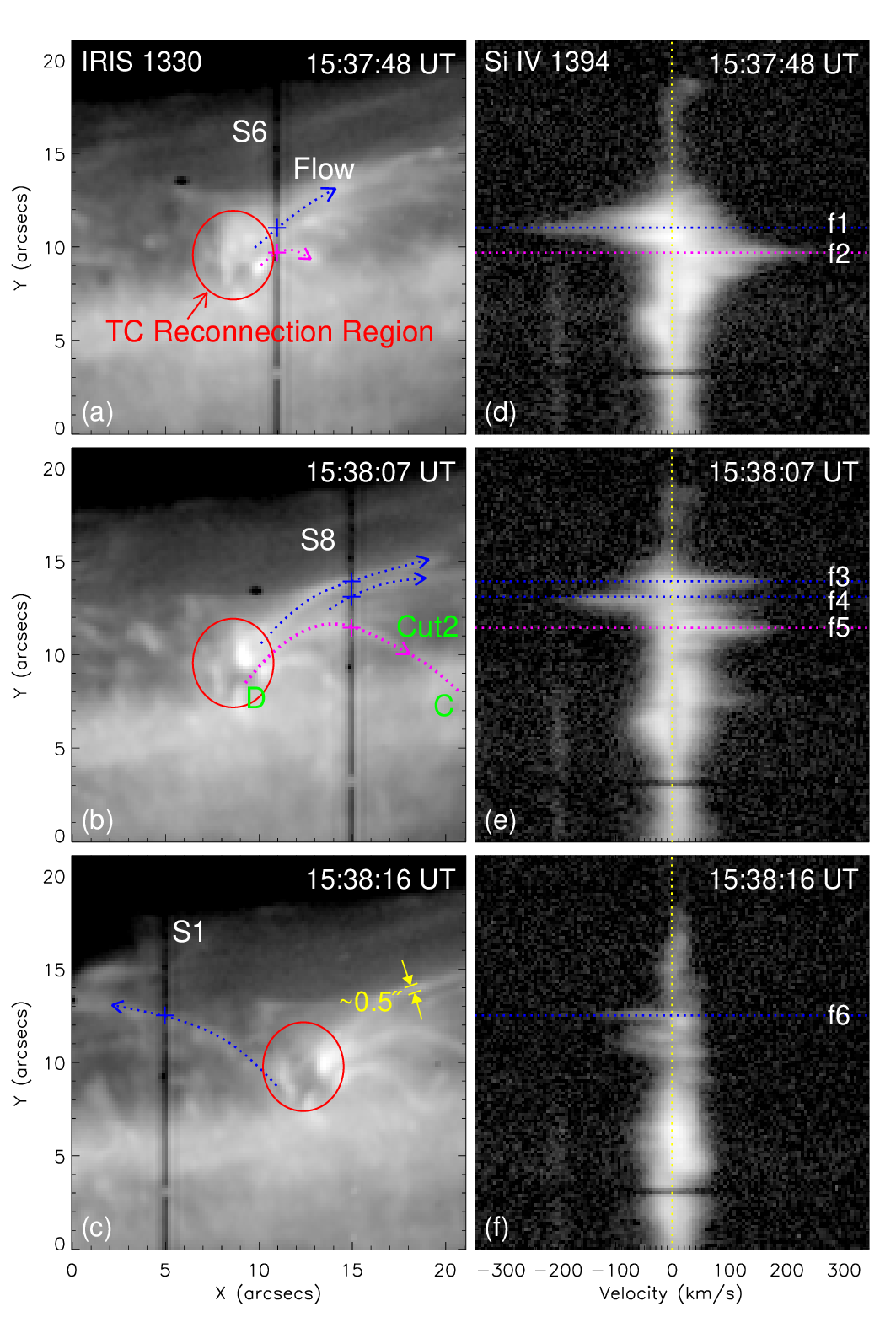}
\caption{((a)--(c)) $IRIS$ 1330 \AA\ SJI images showing the close-ups of the TC reconnection and outflows;
((d)--(f)) the \ion{Si}{4} 1393.76 \AA\ spectra along the slit displayed in panels (a)--(c).
Negative and positive velocities represent wavelengths shorter (blueshift) and longer (redshift) than the rest wavelength, respectively.
The ellipses in panels (a)--(c) indicate the TC reconnection region.
The curved arrows point to the directions of the mass flows along the flux tubes.
The plus signs mark the six positions corresponding to the six features (f1--f6) with strong blueshift or redshift.
The purple dotted curve in panel (b) indicates the cut used to make the time-distance map in Figure~3(h).
\label{fig2}}
\end{figure}


\begin{figure}
\epsscale{0.8}
\plotone{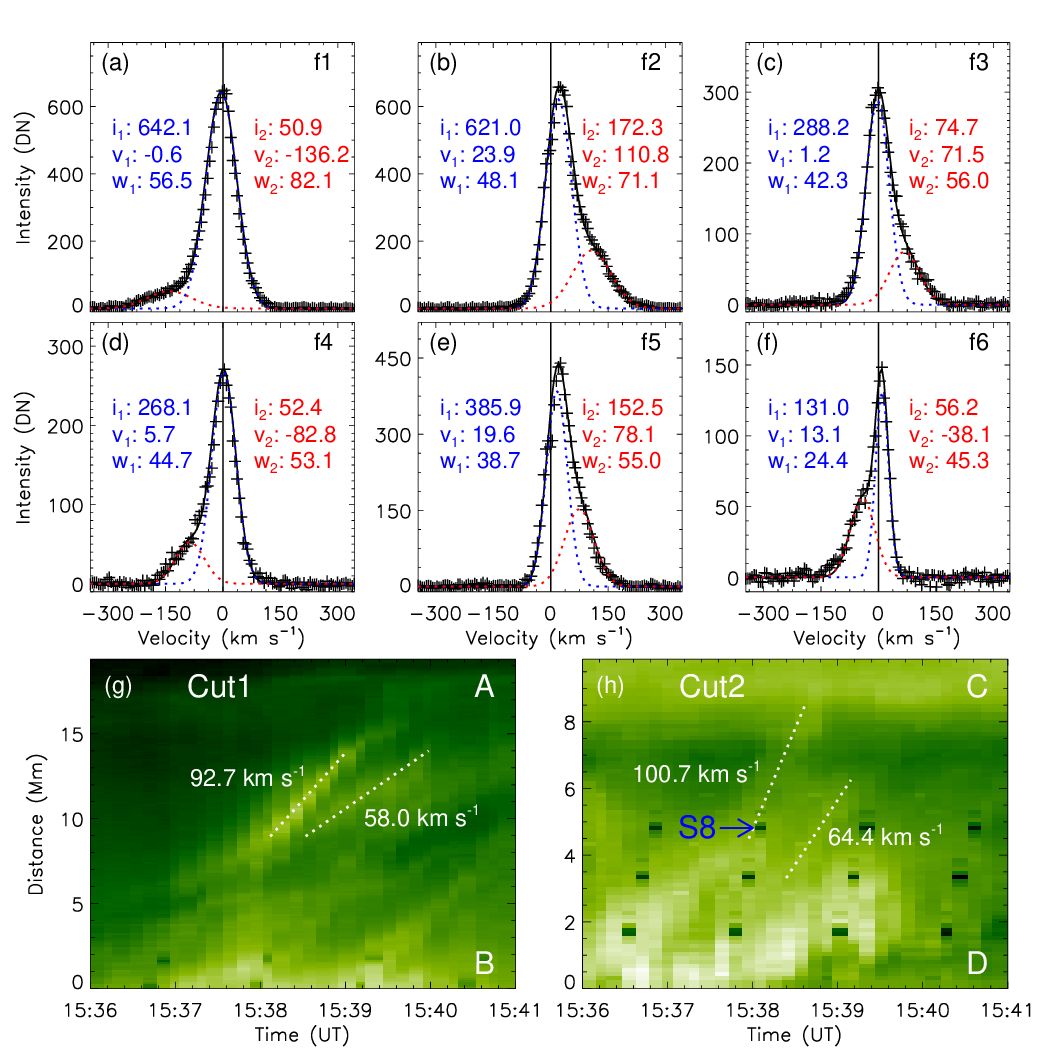}
\caption{((a)--(f)) The \ion{Si}{4} 1393.76 \AA\ line profiles of f1--f6.
The black curves are the double-Gaussian fitting profiles, including two single-Gaussian components which are indicated by the blue and red dotted profiles, respectively.
The peak intensities ($i$; DN), Doppler velocities ($v$; km s$^{-1}$), and line widths ($w$; km s$^{-1}$) of the two components are given by the blue and red values, respectively.
((g)--(h)) Time-distance maps respectively derived from Cut1 and Cut2 as shown in Figures~1(c) and 2(b).
The white dotted lines in panels (g) and (h) indicate the TC reconnection outflows.
\label{fig3}}
\end{figure}

\begin{figure}
\epsscale{0.8}
\plotone{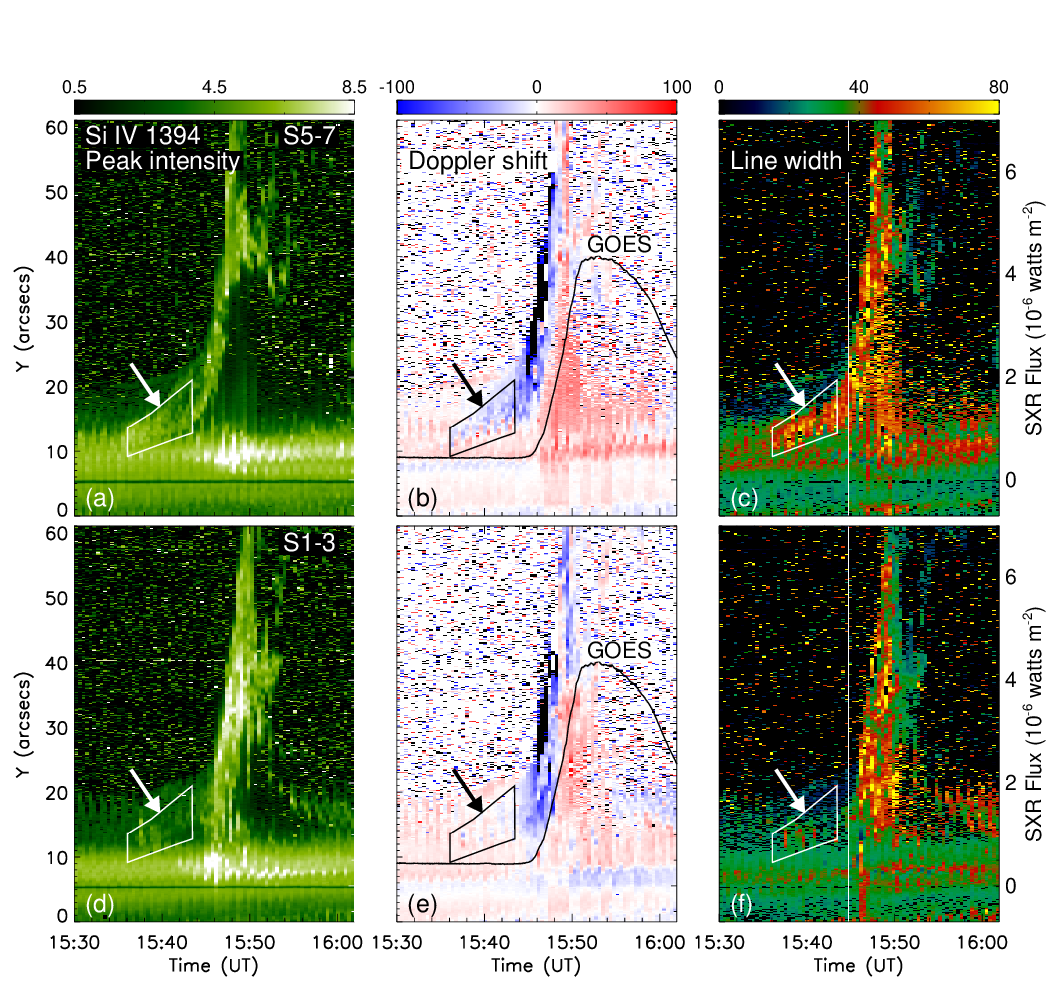}
\caption{Temporal evolution of peak intensity (a), Doppler shift (b), and line width (c) derived from a single-Gaussian fitting to the spectra of \ion{Si}{4} 1393.76 \AA\ line in S5--S7; ((d)--(f)) similar to ((a)--(c)), but for S1--S3.
The spatial ranges are indicated by the discontinuous vertical lines in Figure~1(c).
The black curves in panels (b) and (e) are the $GOES$ SXR light curves.
The arrows and small windows point out the different spectral characteristics between S5--S7 and S1--S3 just before the flare.
\label{fig4}}
\end{figure}

\begin{figure}
\epsscale{1.0}
\plotone{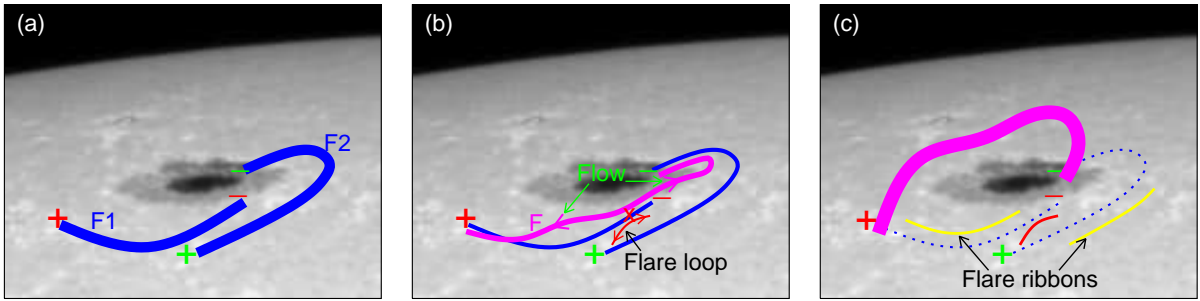}
\caption{Three-dimensional cartoons describing the early phase of the event. 
The backgrounds are an HMI continuum intensity image at 09:09 UT on August 28.
The thinning (thickening) of the blue (purple) curves means the decreasing (increasing) of the field lines of filaments F1 and F2 (erupting filament F).
The FOV is 100\arcsec $\times$ 78\arcsec.
\label{fig5}}
\end{figure}

\end{document}